\documentstyle[sprocl,epsf]{article}

\renewcommand{\bar}[1]{\overline{#1}}
\newcommand{\longvec}[1]{\overrightarrow{\!\!#1}}
\newcommand{\VEV}[1]{\left\langle{#1}\right\rangle}
\newcommand{\lowsim}[1]{\,\mathrel{\rlap{\lower7pt\hbox{$\sim$}}
                    \hskip1pt\hbox{$#1$}}\,}

\newcommand{\etal}{{\em et al.}}

\begin{document}

\title{CP VIOLATION IN  $\tau \rightarrow \nu_\tau+3\pi$}

  \author{YUNG SU TSAI}
\address{Stanford Linear Accelerator Center\\
Stanford University, Stanford, California 94309\\
e-mail: ystth@slac.stanford.edu}

\maketitle\abstracts{
We discuss the ways to find CP violation in the decay of $\tau^\pm
\rightarrow \nu_\tau + 3\pi$
from unpolarized as well as polarized $\tau^\pm$.}

\section{Introduction}

In the Standard Model \cite{1,2} no CP violation can occur in decay
processes involving leptons either as a parent or a daughter because these
processes involve one $W$ exchange and thus, even if the CP violating
complex coupling exists in these decays, its effect will not show up when
the amplitude is squared.  We need two diagrams to interfere with each
other to see the CP violating effects.  The CP violation is caused by the
existence of a complex coupling constant whose phase changes sign as one
goes from particle to antiparticle as required by the hermiticity of
Lagrangian that results in the CPT theorem.  Let $A$ be the complex
coupling constant for $\tau^-$ decay $A=|A|e^{i\delta_t}$, then for
$\tau^+$ decay this coupling constant must be $(CP)\,
A(CP)^{-1}=A^*=|A|e^{-i\delta_t}$ in order to preserve the hermiticity of
the Hamiltonian \cite {3,4}.

In quantum mechanics $i$ is changed into $-i$ under time reversal in order
to preserve the basic commutation relations such as
$[x_i,p_j]=i\delta_{ij}$, $[J_i,J_j]=i\epsilon_{ijk}J_k$, etc.  Thus under
CPT the coupling constant $A$ is
invariant \hfill\break $CPT\, A(CPT)^{-1}=A$.

Let us call the particle exchanged in the new Feynman diagram the $x$
particle.  This particle could be the charged Higgs boson or an entirely
new particle Nature created in order to have CP violation and the
matter-dominated Universe.  In this paper we do not discuss the origin of
such a particle, we restrict ourselves to how the existence of such a
particle with a complex coupling constant can be discovered experimentally.
The $x$ particle must have spin 0 because if it were spin 1 the $x$
exchange diagram must be proportional to the $W$ exchange diagram, thus
even if they have a relative complex phase, the imaginary part of the
relative phase can never show up \cite{3}.  Let $M_x=cM_W$, then
$|M_W+M_x|^2 = |M_W|^2(1+2\, {\rm Re}\, c+|c|^2)$.  Since CP violation is
caused by Im$\,c$, there cannot be any CP violation if $x$ is spin 1.
Higher spin particles are excluded because they are unrenormalizable.

  In $\tau^\pm$ decay CP violation manifests itself in the following way:

\begin{enumerate}

\item
Branching fraction for semileptonic decay of $\tau^-$ with final state
hadronic interactions is different from the corresponding one for $\tau^+$
decay \cite{5}.

\item
The coefficient of $\longvec p_1\cdot\longvec q_i$ in the rest frame of
hadron decay product of $\tau^-$ is different from the corresponding one
for $\tau^+$ lepton, where $\longvec p_1$ is the momentum of $\tau^-$ and
$\longvec q_i$ is the momentum of a decay hadron both measured in the rest
frame of total hadrons cm.  Since $\longvec p_1\cdot \longvec q_i$ is $T$
even, CP violation for this kind of terms can occur only if there is a
strong interaction in the final state because of TCP theorem \cite{3,4,6}.

\item
If $\tau^-$ is polarized with polarization vector $\longvec w$, we have
terms such as $\longvec w\cdot \longvec q_i$, $\longvec w\cdot(\longvec q_i
\times \longvec q_j)$.  Under CP transformation we have
\cite{3,4}
\[
(CP)(\longvec w\cdot\longvec q_i)(CP)^{-1}=-\longvec w^\prime\cdot\longvec
q^\prime_i
\]
and
\[
(CP)(\longvec w\cdot (\longvec q_i\times \longvec q_j))(CP)^{-1}=\longvec
w^\prime\cdot
(\longvec q^\prime_i\times \longvec q^\prime_j) \ .
\]
Hence if the coefficients of these terms in $\tau^+$ decay do not behave as
above then there is CP violation.  Notice that CP changes the sign of
momentum  but does not change the sign of $\longvec w$.

\item
For pure leptonic decay, such as $\tau^\pm \rightarrow
\nu_\tau+\nu_\mu+\mu^\pm$, the only term that violates CP is
$\longvec w_\tau\cdot(\longvec P_\mu\times\longvec w_\mu)$.  Since there is
no final state interaction here we need $T$ odd product to violate CP
because of the CPT theorem.  Hence we need not only $\tau$ polarization but
also measurement of $\mu$ polarization.  This experiment will show whether
a pure leptonic system can have CP violation \cite{7}.
\end{enumerate}

\noindent
Since the direction of polarization can be reversed, it can be used to
isolate the coefficients of different spin dependent terms such as
$\longvec w\cdot\longvec q_i$, $\longvec w\cdot (\longvec q_i\times
\longvec q_j)$.   Thus the polarization is essential for pure leptonic
decay.  For semileptonic decay it is highly desirable especially if the
effect is marginal.

When dealing with CP violation it is of utmost importance to be
open-minded.  The reason is that experimentally\cite {8} there is only
$K_L\rightarrow 2\pi$ known to exhibit CP violation so far and the standard
theory of Kobayashi and
Maskawa\cite{1} was invented to explain this.  It would be a mistake to
neglect testing the CP violation involving leptons just because the CKM
theory says there is no CP violation for $\tau$ decay.  It is also
unconscionable to ignore the possibility of lighter particles participating
in CP violation just because the Higgs exchange model in general favors
heavy particles for this violation!  Fortunately we do not have to assume
any specific model for our purpose.  All we need is to assume the existence
of a spin 0 particle called $x$ that is coupled to leptons and quarks with
CP violating complex coupling constants.  Our task is to devise a method to
test whether the imaginary part of this complex coupling constant is not
zero.  I believe that our first priority is to discover the effect in {\em
any} channel of decay of $\tau$ or semileptonic decay of $B$ or $D$.  After
the effect is discovered in one particular decay mode one can worry about a
systematic search for the $x$ coupling to all quarks and leptons.  Only
after that can one worry about why $x$ exists.

This paper is a sequel to my previous work \cite{4,5,10,11,3} investigating
possible CP violation involving leptons either as a parent or as a
daughter.  There are many issues involved in this vast subject:

\begin{itemize}
\item[(a)]
What is the best decay mode for the first discovery of the effect?
\item[(b)]
The longitudinal polarization of $e^\pm$ helps, but can one do without it?
\item[(c)]
Once the first discovery is made and thus confirming the existence of $x$
boson, we have to
systematically find out the coupling of the $x$ boson to all the leptons
and quarks.  To find out the $x$ coupling to the lepton we not only require
$\tau$ to be polarized but also require the measurement of muon
polarization in the decay $\tau \rightarrow \nu_\mu+\nu_\tau+\mu$.
\item[(d)]
In order to answer the first question (a) we have to theoretically
investigate many decay modes.  This paper deals with one of them; namely,
how to find CP violation in the decay $\tau \rightarrow 3\pi+\nu_\tau$.
\end{itemize}

\noindent
We are not yet in a position to write a comprehensive summary to answer
question (a).  We can give only the
intermediate answer to this question.  For $\tau$ decay:
\begin{enumerate}

\item
The decay mode $\tau^\pm \rightarrow \pi^\pm+\pi^0+\nu$ has the largest
branching fraction (25~\%), but CP violation will be suppressed because it
needs interference between $s$ and $p$ waves of the $2\pi$ produced by $x$
and $W$ bosons.  The $s$ wave is suppressed by the isospin symmetry.  There
is a few \%\ breaking of isosymmetry due to mass difference of $\pi^\pm$
and $\pi^0$, thus the suppression is about the same order of magnitude as
the Cabibbo suppression for decaying into a strange channel \cite{3,4}.

\item
The decay mode $\tau^\pm \rightarrow e^\pm+\nu_e+\nu_\tau$ is next in the
size of branching fraction, but here we need polarization of $\tau^\pm$ as
well as measurement of transverse polarization of $e^\pm$ that is an
impossibly difficult task.

\item
The decay mode $\tau^\pm \rightarrow \mu^\pm+\nu_\mu+\nu_\tau$.  This is
similar to the above, but polarization of $\mu^\pm$ can be measured by its
decay angular distribution.  $\mu^\pm$ can be slow moving if it moves in
the direction opposite to $\tau^\pm$ \cite{5}.

\item
The decay mode $\tau^\pm\rightarrow \pi^\pm\pi^\pm\pi^\mp\nu_\tau$.  This
reaction was first considered by Choi \etal \cite{12}.
In this paper we deal with the same problem with polarization of $\tau$
taken into account.  We also exhibit the phases of $a_1$ and $\pi^\prime$
resonances explicitly, so that it will aid the decision as to how the data
should be integrated to exhibit the existence of the CP violating phase
$\delta_t$.

\item
In Refs. (3) and (11) the Cabibbo suppressed mode $\tau^\pm \rightarrow
K^\pm+\pi^0+\nu_\tau$ is treated.  If $x$ is a charged Higgs boson, this
one could be large.

\item
More complicated cases such as $\tau \rightarrow \nu_\tau +$ more than
$3\pi$, $\tau\rightarrow \nu_\tau+K+$ more than $2\pi$ need to be
considered in the future.

\item
Semileptonic mode without final state interactions such as  $\tau^\pm
\rightarrow \pi^\pm +\nu_\tau$, $\tau^\pm \rightarrow K^\pm + \nu_\tau$
cannot have CP violation because we do not have enough kinematic variables
to construct $T$ odd triple product and the $T$ even term is not allowed
because there is no final state interaction.
\end{enumerate}

For semileptonic decay of hadrons we have:

\begin{enumerate}
\item
$K^\pm \rightarrow \pi^0+\mu^\pm+\nu_\mu$ has been considered \cite{13,14,15}.
\item
$B^\pm \rightarrow \pi^0 + \tau^\pm+\nu_\tau$, $D^\pm \rightarrow
\pi^0+\mu^\pm+\nu_\mu$, $D^\pm \rightarrow
K^0+\mu^\pm+\nu_\mu$, $B^\pm \rightarrow D^0+\tau^\pm+\nu_\tau$ have been
considered  \cite{16,3}.
\item
$t^\pm \rightarrow \pi^0+\tau^\pm+\nu_\tau$, $K^0+\tau^\pm +\nu_\tau$,
$B^0+\tau^\pm +\nu_\tau$ have been considered \cite{17}.
\end{enumerate}

\section{Calculations}

\vspace{.5cm}
\begin{figure}[htb]
\begin{center}
\leavevmode
\epsfbox{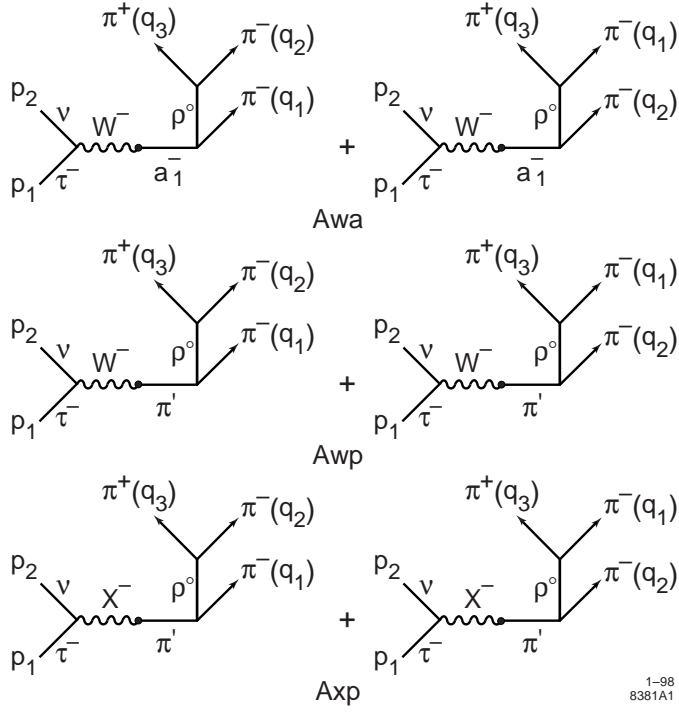}
\end{center}
\caption[*]{Feynman diagrams for $\tau^-\rightarrow
\pi^-+\pi^-+\pi^++\nu_\tau$ for
CP violation.}
\label{fig1}
\end{figure}

We consider six Feynman diagrams shown in Fig. 1 for the decay $\tau^\pm
\rightarrow \pi^\pm +\pi^\mp + \pi^\pm +\nu_\tau$.  There are two identical
particles denoted by $q_1$ and $q_2$.  We may arbitrarily choose which one
is called $q_1$ and $q_2$; for example, $q_1$ and $q_2$ are chosen such
that $s_1 = (q_2+q_3)^2 > s_2=(q_1+q_3)^2$.  In the quark theory both
$a^-_1$ and $\pi^{\prime -}$ consist of  $\bar ud$ combination and thus
they have the same CP violating weak phase $e^{i\delta_w}$ when coupled to
$W^-$ whereas the possible CP violating phase in the $x$ coupling to
$\pi^\prime$ is denoted by $e^{i\delta_x}$.  Only the relative phase
$\delta_t \equiv \delta_x-\delta_w$ will show up in the CP violation.

Let
\begin{eqnarray}
A_{wa} &=& |c_{wa}B_{a_1}|e^{i(\delta_a+\delta_w)}[e^{i\delta_1}M_a(s_1)+
e^{i\delta_2}M_a(s_2)]
\label{1}\\[1ex]
A_{wp} &=&
|c_{wp}B_{\pi^\prime}|e^{i(\delta_p+\delta_w)}[e^{i\delta_1}M_p(s_1)+e^{i\delta_
2}M_p(s_2)]
\label{2}\\[1ex]
A_{xp} &=&
|c_{xp}B_{\pi^\prime}|e^{i(\delta_p+\delta_x)}[e^{i\delta_1}M_p(s_1)+e^{i\delta_
2}M_p(s_2)]
\label{3}
\end{eqnarray}
where
\begin{eqnarray}
c_{wa} &=& \frac{G_F}{2\sqrt 2}\ \cos\theta_cf_a f_{a\rho\pi}
f_{\rho\pi\pi} = |c_{wa}|e^{i\delta_w}
\label{4}\\[1ex]
c_{wp} &=& \frac{G_F}{2\sqrt 2}\ \cos\theta_cf_{\pi^\prime}
f_{\pi^\prime\rho\pi}
f_{\rho\pi\pi} = |c_{wp}|e^{i\delta_w}
\label{5}\\[1ex]
c_{xp} &=& \frac{G_x}{2\sqrt 2}\ f_{\pi^\prime}
f_{\pi\prime\rho\pi}f_{\rho\pi\pi}
= |c_{xp}|e^{i\delta_x}
\label{6}
\end{eqnarray}
$B_{a_1}$, $B_{\pi^\prime}$ and $B_r$ are Breit-Wigner propagators defined by
\begin{equation}B_y(q^2) =
\frac{1}{m^2_y-q^2-im_y\Gamma_y(q^2)}\equiv
\frac{e^{i\delta_y(q^2)}}{\sqrt{(m^2_y-q^2)^2+m^2_y\Gamma^2_y(q^2)}}
\label{7}
\end{equation}
for $y = a_1,\ \pi^\prime,$ or $\rho$.

Since our purpose is to investigate the effect of CP violating phase
$\delta_w$ and $\delta_x$ we explicitly extract all the final state
interaction phases of the problem
$\delta_a,\delta_p,\delta_1$, and $\delta_2$ for $a_1,\ \pi^\prime,\
\rho(s_1)$ and $\rho(s_2)$ respectively.
 \begin{eqnarray}
M_a(s_1) &=&\bar u(p_2)\gamma_\mu(1-\gamma_5)u(p_1)(2q\cdot
q_2q_{3\mu}-2q\cdot q_3q_{2\mu})
|B_r(s_1)| \ , \label{8}\\[1ex]
M_a(s_2) &=& \bar u(p_2)\gamma_\mu(1-\gamma_5)u(p_1)(2q\cdot q_1\cdot
q_{3\mu}-2q\cdot q_3q_{1\mu})
|B_r(s_2)| \ , \label{9}\\[1ex]
M_p(s_1) &=& \bar u(p_2)(1+\gamma_5)u(p_1)(s_2-s_3)|B_r(s_1)|\ ,
\label{10}\\[1ex]
M_p(s_2) &=& \bar u(p_1)(1+\gamma_5)u(p_1)(s_1-s_3)|B_r(s_2)| \ . \label{11}
\end{eqnarray}
The decay rate as well as the decay energy angle distribution of
$\tau^-\rightarrow\nu_\tau +\pi^-+\pi^-+\pi^+$ from a polarized $\tau$ with
a polarization vector $\longvec w$ can be obtained from
\begin{eqnarray}
d\Gamma &=& \frac{1}{2m_\tau}\int \frac{d^3p_2}{2E_2}\ \frac{d^3q_1}{2w_1}\
\frac{d^3q_2}{2w_2}\ \frac{d^3q_3}{2w_3} \ \delta^4(p_1-p_2-q_1-q_2-q_3)
\nonumber \\[1ex]
&&\ \times (2\pi)^{-8}|A_{wa}+A_{wp}+A_{xp}|^2 \ .
\label{12}
\end{eqnarray}

The matrix elements squared can be written as
\begin{eqnarray}
|A_{wa}+A_{wp}+A_{xp}|^2
&=& |A_{wa}|^2+ |A_{wp}+A_{xp}|^2 +
(A^+_{wa}A_{wp}+A^+_{wp}A_{wa})
\nonumber \\[1ex]
&&\  + (A^+_{wa}A_{xp}+A^+_{xp}A_{wa}) \ .
\label{13}
\end{eqnarray}
Using Reduce 3.6 \cite{18} we calculate $|A_{wa}|^2$, $|A_{wp}+A_{xp}|^2$,
 $(A^+_{wa}A_{wp}+A^+_{wp}A_{wa})$ and $(A^+_{wa}A_{xp}+A^+_{xp}A_{wa})$ as
follows:
\begin{eqnarray}
|A_{wa}|^2 &=& 8|c_{wa}|^2|b_a|^2 \left\{ |br_1|^2A_1(q_2)+|br_2|^2A_1(q_1)
\right.
\nonumber\\[1ex]
&&\left.+2|br_1br_2|\left[\cos(\delta_1-\delta_2){\rm Re}\,
A_2-\sin(\delta_1-\delta_2)\,
{\rm Im}\, A_2\right]\right\}
\nonumber\\[1ex]
&&\ + 8|c_{wa}|^2|b_a|^2m_\tau
\nonumber\\
&& \{ w\cdot
q\left[|br_1|^2A_3(q_2)+|br_2|^2A_3(q_1)+2|br_1br_2|\cos(\delta_1-\delta_2)
A_4\right]\nonumber \\[1ex]
&&\ + w\cdot
q_1\left[2|br_2|^2A_3(q_1)+2|br_1br_2|\cos(\delta_1-\delta_2)A_5(q_2)\right]  
\nonumber \\[1ex]
&& \ + w\cdot
q_2\left[2|br_1|^2A_5(q_2)+2|br_1br_2|\cos(\delta_1-\delta_2)A_5(q_1)\right] 
\nonumber \\[1ex]
&&\  + w\cdot
q_3\left[|br_1|^2A_6(q_2)+|br_2|^2A_6(q_1)+2|br_1br_2|\cos(\delta_1-\delta_2)A_7
\right]
\nonumber \\[1ex]
&&\left. - 2|br_1br_2|q\cdot q_3\sin(\delta_1-\delta_2)A_8 \right\}\ ,
\label{14}
\end{eqnarray}
where $b_a=B_{a_1}$, $b_p = B_{\pi^\prime}$, $br_1=B_\rho(s_1)$,
$br_2=B_\rho(s_2)$, $\delta_1$ and $\delta_2$ are phases of $br_1$ and
$br_2$ respectively, and
\begin{eqnarray}
A_1(q_1) &=& 2[(p_1\cdot q_1)(q\cdot q_3)-(p_1\cdot q_3)(q\cdot q_1)]^2
\nonumber \\[1ex]
&& \ + (p_1\cdot q-m^2_\tau)[m^2(q\cdot q_1)^2+m^2(q\cdot q_3)^3
\nonumber \\[1ex]
&&\  -2(q\cdot q_1)(q\cdot q_3)(q_1\cdot q_3)]
\nonumber \\[1ex]
{\rm Re}\, A_2 &=& (p_1\cdot q)[(q\cdot q_1)(q\cdot q_2)m^2-(q\cdot
q_1)(q\cdot q_3)(q_2\cdot q_3)\nonumber \\[1ex]
&&\ -(q\cdot q_2)(q\cdot q_3)(q_1\cdot q_3)  + (q\cdot q_3)^2(q_1\cdot q_2)]
\nonumber \\[1ex]
&&\ + 2.0[(p_1\cdot q_1)(q\cdot q_3)+(p_1\cdot q_3)(q\cdot q_1)]
\nonumber \\[1ex]
&& \ \times[(p_1\cdot q_2)(q\cdot q_3)-(p_1\cdot q_3)(q\cdot q_2)]
\nonumber \\[1ex]
&& \ - (q\cdot q_1)m^2_\tau[(q\cdot q_2)m^2-(q\cdot q_3)(q_2\cdot q_3)]
 \nonumber \\[1ex]
&&\  + (q\cdot q_3)m^2_\tau[(q\cdot q_2)(q\cdot q_3)-(q\cdot q_3)(q_1\cdot
q_2)]
 \nonumber \\[1ex]
{\rm Im}\, A_2 &=& - (q\cdot q_3)q^2Eps(p_1,q_1,q_2,q_3)
\nonumber \\[1ex]
A_3(q_1) &=& 2(q\cdot q_1)(q\cdot q_3)(q_1\cdot q_3)-[(q\cdot q_1)^2 +
(q\cdot q_3)^2]m^2
\nonumber \\[1ex]
A_4 &=& 2.0\Big\{(q\cdot q_3)[(q\cdot q_1)(q_2\cdot q_3)+(q\cdot
q_2)(q_1\cdot q_3)
\nonumber \\[1ex]
&&\ -(q_1\cdot q_2)(q\cdot q_3)] -(q\cdot q_1)(q\cdot q_2)m^2\Big\}
\nonumber \\[1ex]
A_5(q_1) &=& (q\cdot q_3)[-(p_1\cdot q_1)(q\cdot q_3)+(p_1\cdot q_3)(q\cdot
q_1)]
\nonumber \\[1ex]
A_6(q_1) &=& -2.0(q\cdot q_1)[(p_1\cdot q_1)(q\cdot q_3)-(p_1\cdot
q_3)(q\cdot q_1)]
\nonumber \\[1ex]
A_7 &=& (q\cdot q_3)[(p_1\cdot q_1)(q\cdot q_2)+(p_1\cdot q_2)(q\cdot q_1)]
\nonumber \\[1ex]
&&\  - 2(p_1\cdot q_3)(q\cdot q_1)(q\cdot q_2)
\nonumber \\[1ex]
A_8 &=& (q\cdot q_1) Eps (p_2,q_2,q_3,w)-(q\cdot q_2) Eps(p_2,q_1,q_3,w)
\nonumber \\[1ex]
&&\ -(q\cdot q_3) Eps (p_2,q_1,q_2, w)
\nonumber \\[1ex]
m &=& m_\pi = 0.140\ GeV \ , \quad
m_\tau =  1.777\ GeV \ , \quad
q = q_1+q_2+q_3 \ ,
\nonumber \\[1ex]
s_1&=&(q_2+q_3)^2\ ,\quad  s_2=(q_1+q_3)^2\ , \quad s_3=(q_1+q_2)^2 \ .
 \nonumber \\[1ex]
|A_{wp}+A_{xp}|^2 &=&
|bp|^2[|c_{wp}|^2+|c_{xp}|^2+2|c_{wp}c_{xp}|\cos(\delta_x-\delta_w)]
\nonumber \\[1ex]
&& \ \times \Big[M^+_p(s_1)M_p(s_1)+M^+_p(s_2)M_p(s_2)
\nonumber \\[1ex]
&&\ +2M^+_p(s_1)M_p(s_2)\cos(\delta_2-\delta_1)\Big]
\nonumber \\[1ex]
&=& [2.0(m^2_\tau-p_1\cdot q)-2m_\tau w\cdot q]
\nonumber \\[1ex]
&&\ \times [|c_{wp}|^2+|c_{xp}|^2+2c_{wp}c_{xp}\cos
(\delta_x-\delta_w)]\, |bp|^2
\nonumber \\[1ex]
&& \ [|br_1|^2(s_2-s_3)^2+|br_2|^2(s_1-s_3)^2
\nonumber \\[1ex]
&&\ +2|br_1br_2|(s_1-s_3)(s_2-s_3)
\cos(\delta_2-\delta_1)]
\label{15}
\end{eqnarray}
$A^+_{wa}A_{wp}+A^+_{wp}A_{wa}$ can be obtained from $A^+_{wa}
A_{xp}+A^+_{xp}A_{wa}$ by letting $c_{wp} \leftarrow c_{xp}$ and
$\delta_x-\delta_w=0$:
\begin{eqnarray}
 A^+_{wa}W_{wp}+A^+_{wp}A_{wa}&=&  |c_{wa}c_{wp}\, ba\, bp|
\nonumber \\[1ex]
&&\Big\{ 2\cos(\delta_p-\delta_a)\, {\rm Re}\, [M^+_a(s_1)M_p(s_1)]
\nonumber \\[1ex]
&&\ -2\sin(\delta_p-\delta_a)\, {\rm Im}\, [M^+_a(s_1)M_p(s_1)]
\nonumber \\[1ex]
&&\ +2\cos(\delta_p-\delta_a+\delta_2-\delta_1)\, {\rm Re}\,
[M^+_a(s_1)M_p(s_2)]
\nonumber \\[1ex]
&& \ -2\sin(\delta_p-\delta_a+\delta_2-\delta_1)\, {\rm Im}\,
[M^+_a(s_1)M_p(s)2)]
\nonumber \\[1ex]
&& \ + (q_1\leftrightarrow q_1, \ s_1\leftrightarrow s_2)\Big\}\ .
\label{16}
\end{eqnarray}
\begin{eqnarray}
A^+_{wa}A_{xp}+A^+_{xp}A_{wa} &=&
|c_{wa}c_{xp}ba\, bp| \nonumber \\[1ex]
&&\Big\{ 2\cos(\delta_p-\delta_a+\delta_x-\delta_w)\, {\rm Re}\,
[M^+_a(s_1)M_p(s_1)]
\nonumber \\[1ex]
&&\ -2\sin(\delta_p-\delta_a+\delta_x-\delta_w)\, {\rm Im}\,
[M^+_a(s_1)M_p(s_1)] \nonumber \\[1ex]
&&\ +2\cos(\delta_p-\delta_a+\delta_2-\delta_1+\delta_x-\delta_w)\, {\rm Re}\,
[M^+_a(s_1)M_p(s_2)] \nonumber \\[1ex]
&&\ -2\sin(\delta_p-\delta_a+\delta_2-\delta_1+\delta_x-\delta_w)\, {\rm Im}\,
[M^+_a(s_1)M_p(s_2)] \nonumber \\[1ex]
&&\ + (q_1\leftrightarrow q_2,\ s_1\leftrightarrow s_2)\Big\}
\label{17}
\end{eqnarray}
where
\begin{eqnarray}
{\rm Re}\,(M^+_a(s_1)M_p(s_1)) &=& 4m_\tau A_9(q_2) - 4w\cdot q\, A_9(q_2)
\nonumber \\[1ex]
&&\ +4w\cdot q_2(q\cdot q_3)A_{11}(s_2) - 4w\cdot q_3(q\cdot q_2)A_{11}(s_2)
\label{18}
\end{eqnarray}
\begin{equation}
{\rm Im}\,(M^+_a(s_1)M_p(s_1)) = - 4.0\, (s_1-s_3)A_{12}(q_1)
\label{19}
\end{equation}
\begin{eqnarray}
{\rm Re}(M^+_a(s_1)M_p(s_2)) &=& 4m_\tau A_9(q_2) - 4w\cdot qA_{10}(q_1) +
4w\cdot q_1(q\cdot q_3)
A_{11}(s_2) \nonumber \\[1ex]
&&\ + 4w\cdot q_2(q\cdot q_3)A_{11}(s_1)-4w\cdot q_3(q\cdot q_2)A_{11}(s_1) \ .
\label{20}
\end{eqnarray}
\begin{equation}
{\rm Im} M^+_a(s_1)M_p(s_2) = - 4(s_2-s_3)A_{12}(q_1)
\label{21}
\end{equation}
\begin{equation}
A_9(q_2) = (s_2-s_3)[(p_1\cdot q_3)(q\cdot q_2)-(p_1\cdot q_2)(q\cdot q_3)]
\label{22}
\end{equation}
\begin{equation}
A_{10}(q_1) = (s_2-s_3)[(p_1\cdot q_3)(q\cdot q_1)-(p_1\cdot q_1)(q\cdot q_3)]
\label{23}
\end{equation}
\begin{equation}
A_{11}(s_1) = (s_1-s_3)(m^2_\tau-p_1\cdot q)
\label{24}
\end{equation}
\begin{equation}
A_{12}(q_1) = q\cdot q_3\, Eps(p_1,q,q_2,w) - q\cdot q_2\, Eps(p_1,q,q_3,w) \ .
\label{25}
\end{equation}

\subsection{Observations}

\begin{enumerate}
\item
$A_{wa}$ and $A_{wp}$ deal with $W$ exchange, so it deals only with physics
contained in the Standard Model.  The $a_1$ decay amplitude $A_{wa}$ is
more thoroughly investigated than the $\pi^\prime$ decay amplitude $A_{wp}$.

\item
In order to have CP violation we not only need that the gauge bosons
exchanged must be different
($W$ and $x$), but also that the final hadronic states must be different
($a_1$ and $\pi^\prime$).  For example,
Eq. (\ref{15}) shows that the interference between $A_{wp}$ and $A_{xp}$ is
proportional to $\cos(\delta_x-\delta_w)$ for $\tau^-$ decay whereas for
$\tau^+$ decay it is $\cos(\delta_w-\delta_x)$ which is equal to
$\cos(\delta_x-\delta_w)$ and thus we cannot have a CP violating effect
even if $\delta_w-\delta_x\ne 0$.  This is because $A_{wp}$ and $A_{xp}$
are proportional to each other and thus the imaginary part of the phase
difference does not show up.  Equation (\ref{16})  does not contain
$\delta_w$ or $\delta_x$, so no CP violating effect.  Only Eq. (\ref{17})
can contain CP violating effects.

\item
When $\tau$'s are not polarized only $A_9(q_1)$ and $A_9(q_2)$, shown in
Eq. (\ref{22}), appear
in the CP violating expression.  CP violation can thus be checked by
comparing the coefficients of
$A_9(q_1)$ and $A_9(q_2)$ for $\tau^-$ decay with those of
$A_9(q^\prime_1)$ and $A_9(q^\prime_2)$ for $\tau^+$ decay.  If they are
different, then CP is violated.

\item
We are interested in the CP violation in the {\em decay} of $\tau$.  The CP
violation in the {\em production} of $\tau$ is expected to be much less
than $\alpha/\pi$ that is caused mainly by the possible existence of the
electric dipole moment of tau \cite{19}.  If we ignore the CP violation in
the production then \hbox{\boldmath$w=w^\prime$}; namely, the polarization
vectors of $\tau^+$ and $\tau^-$ must be equal and parallel to each other
in the center-of-mass system\cite{4} of $e^\pm$.  When the energy is near
threshold such as in the case of the Tau-Charm Factory we have $s$ wave
production.  In the $s$ wave production we have \cite{4}
\begin{equation}
\mbox{\boldmath$w$} = \mbox{\boldmath$w^\prime$}
= \widehat e_z\ \frac{w_1+w_2}{1+w_1w_2}\ ,
\label{26}
\end{equation}
where $w_1$ and $w_2$ are longitudinal polarization of $e^-$ and $e^+$ in
the incident $e^-$ direction that is chosen as the $z$ axis.  For the $B$
factory energy \hbox{\boldmath$w$} is given in Ref. (4).

\item
From Eqs. (\ref{17})-(\ref{25}), we see that there are much more
polarization dependent terms than the polarization independent terms.  Thus
the polarization gives us more handles to find out whether CP is violated
or how it is violated.  For example, in Eq. (\ref{17}) the sine terms
involve only $A_{12}$ given by Eq. (\ref{25}) which vanishes when
\hbox{\boldmath$w$} $=0$.  The sign of $\tau^\pm$ polarization can be
reversed by reversing the polarization of initial $(e^-+e^+)$.  This will
help in isolating different polarization dependent terms such as the
coefficients of $\longvec w\cdot \longvec q_1$, $\longvec w\cdot \longvec
q_2$, $\longvec w\cdot(\longvec q_1\times\longvec q_3)$, etc.
\end{enumerate}

\section{Conclusions}

This is a series of papers investigating new mechanisms of nonstandard CP
violation using $\tau$ decay and semileptonic decay of a hadron.  If CP
violation is discovered in any one of these decays it will imply the
existence of a spin zero boson $x$ transmitting a CP nonconserving force.
After the discovery in any one channel we can then systematically
investigate how $x$ is coupled to all particles.  This will enable us to
construct a correct theory of CP violation.  The attractive feature of this
mechanism is that it will allow leptons to participate fully in CP
violation.  The standard theory is unnatural in the sense that it will not
allow leptons to participate in CP violation.

I am honored to be able to present this paper in memory of Professor
C.~S.~Wu who first used angular distribution of a decay lepton from a
polarized cobalt nucleus to discover the existence of
$\VEV{\longvec\sigma}_{\rm cobalt}\cdot\longvec p_e$ term and thus the
existence of the parity violation.  The phenomenon discussed in this paper
is a natural extension of her work in the sense that in the $e^\pm$
colliding machine particle and antiparticle parents $(\tau^\pm)$ are
produced equally abundantly.  By comparing the decay angular distributions
of polarized or unpolarized $\tau^\pm$ we hope to be able to discover a new
mechanism of CP violation.  I thank
Professor Fang Wang for inviting me to the memorial conference.  Numerical
work for this paper is still in progress.
The difference in detection efficiency between particle and antiparticle
can be resolved by a combination of experimental and theoretical means.
Experimetally the decay modes $\tau \rightarrow \pi +\nu$ and $\tau
\rightarrow K+ \nu$ do not have final state interactions and hence their
branching fractions as well as the energy-angle distributions are not
affected by the CP violations. Thus these decay modes can be used to
experimentally check the difference in detection efficiency between
particle and antiparticles. The difference in detection efficiency comes
from difference in cross sections between particle and antiparticles on the
detector materials and they can be estimated theoretically. This
calculation is in progress.

The difference in detection efficiency between particle and antiparticles
can be resolved by a combination of experimental and theoretical means.
Experimetally the decay modes $\tau \rightarrow \pi +\nu$ and $\tau
\rightarrow K+ \nu$ do not have final state interactions and hence their
branching fractions as well as the energy-angle distributions are not
affected by the CP violations. Thus these decay modes can be used to
experimentally check the difference in detection efficiency between
particle and antiparticles. The difference in detection efficiency comes
from the difference in cross sections between particle and antiparticles on the
detector materials and they can be estimated theoretically. This
calculation is in progress.

\end{document}